# Flavor–Parity Breaking in the NJL Model with Wilson Fermions.

Khalil M. Bitar,[a]  Pavlos M. Vranas, [*] [b]

[a]SCRI, The Florida State University, Tallahassee, FL 32306-4052, U.S.A.

[b]Physics Department, Columbia University, New York, NY 10027, U.S.A.

The Nambu–Jona-Lasinio model is chirally symmetric. Addition of a Wilson term explicitly breaks this symmetry and leaves the model with a remaining parity-flavor symmetry. In the approximation of a large number of colors it has been shown that there is a phase where the remaining parity-flavor symmetry is spontaneously broken and that on the phase boundary all pions become massless. Using numerical simulations we confirm the existence of this phase for the case of two flavors and two colors. We also confirm the large N prediction of the existence of large finite size effects that alter the shape of the phase boundary dramatically when periodic boundary conditions are used.

The lattice Nambu–Jona-Lasinio model is an interesting toy model of lattice QCD. For an extensive analysis see [1] and references therein. The Lagrangian of the two flavor model in Minkowski space and in continuum notation is:

$$\mathcal{L} = \overline{\Psi}(i\not\partial - m_0)\Psi + \frac{G_1}{2}\left[(\overline{\Psi}\Psi)^2 + (\overline{\Psi}i\gamma_5\boldsymbol{\tau}\Psi)^2\right]. \quad (1)$$

The fermionic field $\Psi$ is a flavor $SU(2)$ doublet and a color $SU(N)$ vector. $\boldsymbol{\tau} = \{\tau_1, \tau_2, \tau_3\}$ are the three isospin Pauli matrices, $\not\partial = \gamma^\mu \partial_\mu$, and $m_0$ is the bare quark mass. The Lagrangian is diagonal in color, in contrast with full QCD where it is diagonal in flavor. The partition function is obtained after rotation to Euclidean space, addition of a Wilson term and introduction of scalar $\sigma$ and pseudoscalar $\boldsymbol{\pi}$ auxiliary fields (to make the action quadratic in the fermionic fields) [1]. The Wilson term breaks the chiral symmetry explicitly down to flavor symmetry. Therefore, if there are massless pions they can not be the Goldstones of a spontaneously broken chiral symmetry. What is then the mechanism that provides massless pions? This question was posed in [2] for QCD and recently in [3] for the NJL model. There the NJL model is analyzed using large $N$ methods. The saddle point equations are:

$$\sigma_s \frac{\beta_1}{2N} - \int \frac{d^4p}{(2\pi)^4} \frac{\sigma_s + m_0 + rw(p)}{g(p, m_q)} = 0$$

[*]Speaker

$$\pi_s \frac{\beta_1}{2N} - \int \frac{d^4p}{(2\pi)^4} \frac{\pi_s}{g(p, m_q)} = 0 \quad (2)$$

with

$$g(p, m_q) = \sum_\nu \sin^2 p_\nu + [rw(p) + m_q]^2$$

$$w(p) = 4 - \sum_\mu \cos p_\mu, \quad m_q = m_0 + \sigma_s \quad (3)$$

where $\sigma_s, \pi_s = |\boldsymbol{\pi}_s|$ are the uniform saddle point fields, $\beta_1 = \frac{1}{4G_1}$, $r$ is the coefficient of the Wilson term (we will set $r = 1$), and $m_q$ is the constituent quark mass. There are two solutions to these equations corresponding to two phases. (**I**) A flavor-parity symmetric phase where the order parameter $\boldsymbol{\pi}_s = 0$ ($\leadsto (\overline{\Psi}i\gamma_5\boldsymbol{\tau}\Psi)_s = 0$) and massive pions $m_{\pi_1}, m_{\pi_2}, m_{\pi_3} \neq 0$. (**II**) A spontaneously broken flavor-parity symmetry phase where, for example, $\boldsymbol{\pi}_s = (0, 0, \pi_s)$, $\pi_s \neq 0$ ($\leadsto (\overline{\Psi}i\gamma_5\boldsymbol{\tau}\Psi)_s \neq 0$), two pions are massless $m_{\pi_1} = m_{\pi_2} = 0$, and one is massive $m_{\pi_3} \neq 0$. The two massless pions are the Goldstones of the spontaneously broken flavor symmetry. The mass of the third pion can be interpreted as the inverse correlation length associated with the spontaneous breaking of parity symmetry. On the phase boundary this discrete symmetry undergoes a second order Ising-like phase transition where the associated correlation length diverges and therefore the mass of the third pion vanishes. So the phase boundary, which is a line in the $\beta_1, m_0$ plane, is characterized by $\pi_s = 0$, $m_{\pi_1} = m_{\pi_2} = m_{\pi_3} = 0$. This phase diagram, determined from the saddle point equations, was calculated in [3] for an infinite volume lattice using a Newton procedure to perform the integrals. Here we will concentrate our efforts in finite size lattices for two reasons. First, we will perform numerical simulations for finite $N = 2$ in order to confirm the existence of the spontaneously broken phase. Second, on finite lattices the zero fermionic modes alter the shape of the phase diagram significantly. We study this using large $N$ methods and numerical simulations.

On a finite lattice the phase diagram can be computed from the saddle point equations at the transition point ($\pi_s = 0$):

$$\frac{m_0}{L_x^3 L_t} \sum_k \frac{1}{g(k, m_q)} + \frac{r}{L_x^3 L_t} \sum_k \frac{w(k)}{g(k, m_q)} = 0 \quad (4)$$

$$\beta_1 - \frac{2N}{m_q - m_0} \frac{1}{L_x^3 L_t} \sum_p \frac{m_q + rw(p)}{g(p, m_q)} = 0 \quad (5)$$

For given $m_q$ we calculate $m_0$ from 4. Inserting this into 5 we find $\beta_1$. On a periodic lattice the momentum takes values $p_i = \frac{2\pi n_i}{L_i}, n_i = 0, 1, \cdots L_i - 1$. Then $g(p, m_q)$ vanishes at:
$m_q = 0, \quad p = (0, 0, 0, 0)$,
$m_q = -2, \quad p = (\pi, 0, 0, 0)$ and permutations,
$m_q = -4, \quad p = (\pi, \pi, 0, 0)$ and permutations,
$m_q = -6, \quad p = (\pi, \pi, \pi, 0)$ and permutations,
$m_q = -8, \quad p = (\pi, \pi, \pi, \pi)$.
The 16 momenta where $g(p, m_q)$ vanishes correspond to the origins of the Brillouin zones of the 16 doubler species. As a result, at these values of $m_q$, the corresponding terms of the momentum sum in eq. 5 are singular, and therefore so is the function $\beta_1(m_0)$ that describes the phase boundary. This is shown for an $8^3 \times 16$ size lattice by the solid line in Figure 1.

Inside the octopus-like graph the flavor-parity symmetry is spontaneously broken. At point "1" of this figure $m_q$ is very large and positive. As $m_q$ decreases, crosses zero and tends to very large negative values, we transverse the whole solid line from "1" to "2" to "3" $\cdots$ to "12" where $m_q$ is very large and negative. The "prongs" of this figure, points "3", "5", "7", "9", and "11", extend all the way to $\beta_1 = \infty$ where they correspond to $m_q = 0, -2, -4, -6, and -8$, respectively. When the singular terms described above are neglected from the calculation (the contribution of these terms disappears in the infinite volume limit) we obtain the dotted line of Figure 1. This line is quantitatively very close to the infinite volume phase line of [3]. As the volume increases to infinity the "spikes" of the solid line shrink in "thickness" and eventually are mapped close to the dotted line, producing the infinite volume phase line of [3]. These spikes are a very strong finite size effect. For example, at finite volume, in order to obtain a very small $m_q$, one has to be near point "3", i.e. at very large $\beta_1$. As the volume becomes infinite very small values of $m_q$ can be obtained at finite $\beta_1$ close to point "2". For more details and figures see the second reference in [1].

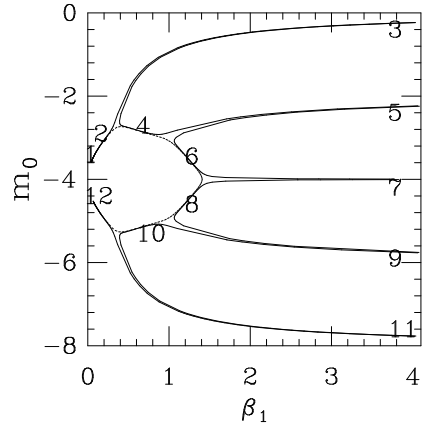

**Figure 1.** Phase diagram with periodic BC

When antiperiodic BC are used even along one direction (say along the time direction) $g(p, m_q)$ has no zeros (for example the smallest momentum is $(0, 0, 0, \frac{\pi}{L_t})$). As a result, $\beta_1(m_0)$ has no singularities. The spikes are still present but they do not extend to $\beta_1 = \infty$. Their extension along the $\beta_1$ direction is roughly proportional to $\frac{L_t}{L_x L_y L_z}$. The phase diagram for lattice sizes $16^3 \times 32$, $16^3 \times 64$, $8^3 \times 16$, $8^3 \times 64$ is shown in Figure 2. In the order mentioned above the $16^3 \times 32$ has the smallest size "spikes" and the $8^3 \times 64$ the largest.

In order to confirm the large $N$ predictions we have performed Hybrid Monte Carlo simulations on the CM-2 supercomputer at SCRI on $8^3 \times 16$

lattices at $N = 2$. In the parity broken phase we define the "magnetization"

$$M_\pi = \sum_{x,t} \pi_{x,t}, \qquad \pi_{x,t} = i\overline{\Psi}_{x,t}\gamma_5\tau\Psi_{x,t} \ . \qquad (6)$$

The massive pion propagator is measured from the decay in the "t" direction of the correlation function $c(t)$ of the component of $\pi$ along $M_\pi$.

$$c(t) = <\sum_x [M_\pi \cdot \pi_{x,t}][M_\pi \cdot \pi_{o,o}]> \qquad (7)$$

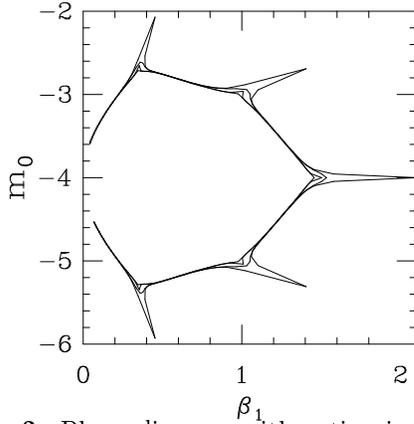

Figure 2. Phase diagram with antiperiodic BC along the time direction

The results for the pion mass at $\beta_1 = 2.5$ are denoted by crosses in Figure 3. The solid line is the large $N$ prediction. This confirms both the existence of a spontaneously broken flavor-parity phase as well as the presence of a "spike" at that region of the phase diagram. The same plot, but for $\beta_1 = 0.5$, is shown in Figure 4. In this case, as can be seen from Figure 1, when $m_0$ is varied from $-3$ to $-1.5$ one crosses the phase line three times. The crossing at $m_0 \approx -2.76$ is predicted fairly well by the numerical results. The other two crossings entail a very small pion mass and are missed by the numerical results (albeit some indication from the shape of the curve). Similar conclusions can be drawn by looking at the plot of $|M_\pi| = M_\pi$ in Figure 5.

In conclusion, both large $N$ predictions, namely the existence of a spontaneously broken flavor-parity phase as well as the presence of the "spikes" on finite lattices with periodic boundary conditions, were verified numerically at $N = 2$.

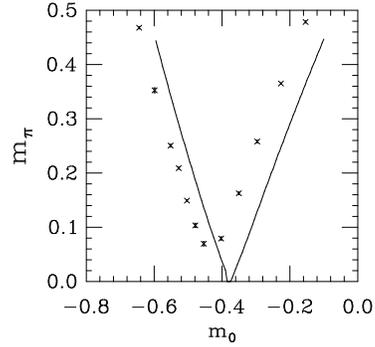

Figure 3. $\beta_1 = 2.5$, lattice size is $8^3 \times 16$

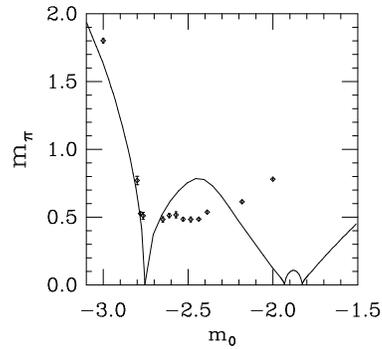

Figure 4. $\beta_1 = 0.5$, lattice size is $8^3 \times 16$

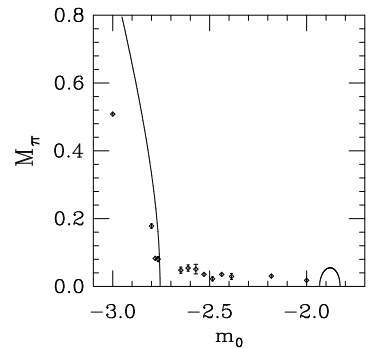

Figure 5. $\beta_1 = 0.5$, lattice size is $8^3 \times 16$

This research was supported by the DOE under grant DE-FG05-92ER40742 for KB and PV, and also under grant DE-FG02-92ER40699 for PV.